\documentclass[aps,twocolumn,showpacs]{revtex4}
\usepackage{epsfig,amsmath,psfrag,graphicx,dsfont,mathrsfs,subfigure}

\newcommand\ket[1]{\left|#1\right\rangle}
\newcommand\bra[1]{\left\langle#1\right|} 

\newcommand\ketbra[2]{ | #1 \rangle\langle #2 | }

\newcommand\expect[1]{\langle#1\rangle}
\newcommand\average[1]{\left\langle\!\left\langle#1\right\rangle\!\right\rangle}
 
\newcommand{\eps}{\mathcal E} \newcommand{\id}{\mathds 1}
\newcommand{\ten}{\otimes} \newcommand{\sigz}{\sigma_z}
\newcommand{\tr}{\mbox{tr}} \renewcommand{\rho}{\varrho}

\newcommand{\Cd}{C_{\mbox{\tiny D}}}
\newcommand{\Cw}{C_{\mbox{\tiny W}}}
\newcommand{\Cm}{C_{\mbox{\tiny MEMS}}}
\newcommand{\Cmax}{C_{\mbox{\tiny Max}}}

\begin{document}

\title{Decoherence and Entanglement Dynamics in Fluctuating Fields}
\author{Julius Helm}
\affiliation{Institut f\"{u}r Theoretische Physik, Technische Universit\"{a}t Dresden, 
D-01062 Dresden, Germany}
\author{Walter T. Strunz}
\affiliation{Institut f\"{u}r Theoretische Physik, Technische Universit\"{a}t Dresden, 
D-01062 Dresden, Germany}

\date{\today}
\begin{abstract}
  We study pure phase damping of two qubits due to fluctuating
  fields. As frequently employed, decoherence is thus described in
  terms of random unitary (RU) dynamics, that is, a convex mixture of
  unitary transformations. Based on a separation of the dynamics into
  an average Hamiltonian and a noise channel, we are able to
  analytically determine the evolution of both entanglement and
  purity. This enables us to characterize the dynamics in a
  concurrence-purity (CP) diagram: We find that RU phase-damping
  dynamics sets constraints on accessible regions in the CP plane. We
  show that initial state and dynamics contribute to final
  entanglement independently.
\end{abstract}
\pacs{03.65.Yz,03.65.Ud,03.67.Pp}
\maketitle

\section{Introduction}

For newly emerging quantum technologies, robustness of quantum states
is essential. In particular, stability of entanglement spread over a
many-body quantum system is of fundamental importance for quantum
cryptography and quantum information processing in general. A thorough
understanding of processes known to destroy the desired quantum
qualities---usually subsumed under the name of decoherence---is
necessary. Phase damping represents the case of pure decoherence,
where coherences of a given state in a certain basis are subject to
decay while probabilities, that is diagonal elements of the density
matrix, remain unchanged. It is well known that despite the rather
simple nature of phase damping, it is clearly enough to disentangle
quantum states \cite{Yu2003, Mintert2005}.

A widely encountered source of decoherence is growing entanglement
between a quantum system and its environment. Yet, often decoherence
is due to---or at least may be explained in terms of---random unitary
(RU) dynamics. Then, fluctuating classical fields are liable for the
loss of quantum properties (sometimes also termed ``random external
fields'') \cite{AlickiLendi, NielsenChuang}. Although it is known that
RU dynamics is not the most general form of decoherence
\cite{LandauStreater, HelmStrunz}, it is of great practical
importance. For instance, in quantum computers based on trapped ions,
these classical fluctuations are believed to be the main source of
decoherence \cite{Monz2009}. Here, fluctuations are present both in
the magnetic field of the trap and in the frequency of the laser
addressing the qubits. RU dynamics moreover serves as a model to
introduce decoherence in experiments in a controlled fashion
\cite{Myatt2000}. In the context of quantum error correction, RU
processes are of special interest, because they represent the only
type of errors that may be completely undone \cite{GregorattiWerner}.
The present work will further reveal how decoherence of qubits due to
RU dynamics allows for a surprisingly far-reaching analytical
treatment.

For a system of two qubits a possible means of characterization relies
upon the relation between entanglement and entropy \cite{Ishizaka2000,
  Ziman2005}. The system's state $\rho$ may be studied in terms of a
CP diagram, where entanglement (here measured through concurrence $C$)
is plotted against purity, $P(\rho) = \tr (\rho^2)$, a common measure
for the mixedness of $\rho$. This approach has been used in several
theoretical studies, where Bell states subject to single-sided
dynamics \cite{Ziman2005} (by {\it single-sided dynamics} we denote
the case where only one part of a bipartite system is affected),
decoherence in a random matrix environment \cite{Pineda2007}, or two
interacting atoms in a cavity \cite{Torres2009} were considered.

It was noted earlier that decoherence in ion trap quantum computers is
not of simple exponential type \cite{Schmidt-Kaler2003}, as would be
expected from a Markovian master equation approach. A theoretical
model incorporating this fact thus calls for a more general
treatment. From an axiomatic point of view (neglecting initial
correlations) the dynamics of a quantum system is given in terms of
completely positive maps (or {\it quantum channels})
\cite{NielsenChuang}. In a Hilbert space of dimension $d$, these
channels can always be written in terms of at most $d^2$ {\it Kraus
  operators} $K_i$ such that
\begin{equation}
  \label{eq:kraus}
  \rho \mapsto \rho' = \eps[\rho] = \sum_i K_i \rho K_i^\dagger
\end{equation}
(throughout the article we denote the initial state by $\rho$ and its
map by $\rho'$). For RU channels there exists an expression of the
form (\ref{eq:kraus}) where every Kraus operator is proportional to
some unitary operator, so that the dynamics is a convex combination of
unitary transformations: $\rho' = \sum_i p_i U_i \rho U_i^\dagger$,
with $p_i>0$ and $\sum_i p_i = 1$. Phase-damping channels stand out
due to the requirement to be diagonal in the distinguished basis, that
is, they may be written in the form
\begin{equation}
  \label{phase dampingchannel}
  \rho'_{mn} = \langle a_n|a_m\rangle \rho_{mn} =: D_{mn} \rho_{mn},
\end{equation}
with $\{ \ket{a_n} \}$ any set of $d$ normalized complex vectors
\cite{Havel,Paulsen}. In ``system plus environment'' models,
the vectors $|a_n\rangle$ may be identified with environmental
quantum states \cite{GorinStrunz2004}. For RU phase damping,
a simple interpretation of the $|a_n\rangle$ is less obvious.

In this article we study RU phase damping of two qubits. A fully
analytical expression for both the entanglement evolution and the
purity decay is obtained. Throughout the article the initial two-qubit
state is assumed to be pure. The phase damping will be realized as an
ensemble average $\average{U(t)\rho U^\dagger(t)}$, where the unitary
(diagonal) time evolution may be regarded as arising from a stochastic
Hamiltonian of the form
\begin{eqnarray}
  \label{eq:hamil-zweiqubit}
  H(t) = 
  \omega_1(t) \sigz^{A}\ten \id + \omega_2(t) \id \ten \sigz^{B} 
  + \omega_3(t) \sigz^A \ten \sigz^B. 
\end{eqnarray}
With $\sigz$ we denote the diagonal Pauli spin operators for qubits
$A$ and $B$, respectively, the time dependence is due to the
stochastic processes $\omega_1(t), \omega_2(t)$, and $\omega_3(t)$. In
an experimental realization using two-state atoms this would
correspond, for example, to a fluctuation of the individual Zeeman
levels described by $\omega_1(t)$ and $\omega_2(t)$, together with an
instability in the interaction of the atom's energy eigenstates,
described by $\omega_3(t)$. Such a Hamiltonian is used to describe the
spin-spin interaction in nuclear magnetic resonance (NMR) systems
\cite{Vandersypen2004}. Also note that the third term of
Eq.~(\ref{eq:hamil-zweiqubit}) describing the interaction---later
referred to as pure two-qubit phase damping---is implemented in the
realization of a phase gate in recent ion trap experiments
\cite{Monz2009}.

Recently, for a quantum system subject to single-sided dynamics a very
simple evolution equation for entanglement was found
\cite{Konrad2007}.  This evolution equation allows for a factorization
into a functional characterizing the quantum channel and the amount of
the initial pure-state entanglement. In our case here, the evolution
will in general be nonlocal. Yet, we are also able to give some exact
evolution equations for entanglement under phase damping, where the
channel and the initial state enter independently.

The article is organized as follows: In Sec. II we discuss our model
of RU phase damping, where we benefit from a separation of the channel
into a reversible part due to a mean Hamiltonian and an irreversible
noise channel. This separation enables us to analytically study
changes in entanglement and purity. In Sec. III we discuss RU
single-qubit channels in the context of the CP diagram. Here we are
able to give an alternative explanation of the results obtained in
Ref.~\cite{Ziman2005}, where for a maximally entangled state subject
to unital (mapping the completely mixed state onto itself)
single-sided channels certain bounds to the accessible area within the
CP diagram were found. Our approach, which employs the so-called
Jamiolkowski isomorphism, will also be well suited for explaining part
of the findings at hand. In Secs. IV and V we analyze in detail the
evolution of entanglement and purity of a pure separable and a general
pure initial state under RU phase damping, respectively.

\section{RU Phase Damping}
\label{sec:RUPD}

The phase damping we consider shall be realized as an ensemble average
$\rho' = \average{U(t)\rho U^\dagger (t)}$, where we may set $U(t) =
e^{-i \int_0^t d\tau H (\tau)}$, a unitary map based on a stochastic
diagonal Hamiltonian $H(t)$. As an example let us consider a
single-qubit channel, where a generic diagonal Hamiltonian may simply
be set to $H(t) =\omega(t) \sigz$ with some stochastic process
$\omega(t)$. Assuming $\omega(t)$ to add up to a total perturbation of
the qubit, the central limit theorem tells us that $\Omega := \int_0^t
d\tau \omega(\tau)$ is a Gaussian random variable that is determined
by its mean, $\mu:=\average{\Omega}$, and its variance,
$\sigma^2=:\average{ \Omega^2}-\mu^2$, hence leading to $\average{
  e^{\pm i \Omega t}}= e^{\pm i\mu} e^{-\frac{1}{2} \sigma^2}$.

Substituting $\mu \rightarrow \theta$ and $\frac{1}{2}
(1-e^{-\frac{1}{2} \sigma^2}) \rightarrow p$, we thus arrive at the
most general form of a single-qubit phase-damping channel
\begin{eqnarray*}
  \rho' &=& 
 \begin{pmatrix}
   \rho_{11} & \average{e^{-i \Omega t}} \rho_{12} \\
   \average{e^{i \Omega t}}\rho_{21} & \rho_{22}
 \end{pmatrix} \\
 &=& e^{-i\frac{\theta}{2}\sigma_z} \left((1-p) \; \rho + p \;
   \sigma_z \rho \sigma_z\right) e^{i\frac{\theta}{2}\sigma_z}.
\end{eqnarray*}
Thus, single-qubit phase damping may always be written in terms of RU
dynamics \cite{LandauStreater}. From our example we see furthermore
that it suffices to consider Gaussian fields.

For two qubits the corresponding stochastic diagonal Hamiltonian is
\begin{eqnarray*}
  H(t) = 
  \omega_1(t) \sigz^A\ten\id + \omega_2(t) \id \ten\sigz^B 
  + \omega_3(t) \sigz^A\ten \sigz^B. 
\end{eqnarray*}
Irrespective of possible correlations among the accumulated phases
$\int_0^t d\tau \omega_k(\tau) =: \Omega_k$, we find that the
phase-damping channel can be decomposed according to
\begin{eqnarray}
  \label{eq:commutativity}
  \rho' &=& U_\mu^{} (\tilde D \star \rho) U_\mu^\dagger \\
  &=& \tilde D \star \left(U_\mu^{} \rho U_\mu^\dagger\right) \nonumber
\end{eqnarray}
into a unitary, reversible part $U_\mu$ based on the mean Hamiltonian
$H_\mu = \mu_1 \sigz^A\ten\id + \mu_2 \id\ten\sigz^B + \mu_3
\sigz^A\ten\sigz^B$ ($\mu_k:=\average{\Omega_k}$) and a nonunitary,
irreversible noise channel $\tilde D$ (see Appendix
\ref{app:RUchannels} for more details). For brevity of notation we
here made use of the {\it Hadamard} poduct $\star$ of matrices, that
is, the pointwise multiplication of two matrices of the same size
\cite{HornJohnson}.

In close analogy to the single-qubit case we can further deploy the RU
phase-damping channel by assuming $\vec \Omega$ to be a Gaussian
process. It should be noted, however, that our analysis could easily
be extended to more general statistics. We can then use the
characteristic function \cite{Honerkamp}
\begin{eqnarray*}
  \average{e^{\pm i \vec k \vec \Omega}} = e^{-\frac{1}{2} \vec k^T \Sigma
    \vec k \, \pm \,  i \vec k \vec \mu } \quad \forall \vec k \in \mathds C^3, 
\end{eqnarray*}
where $\vec \mu := \langle\!\langle \vec\Omega \rangle\!\rangle $
denotes the mean value, $\Sigma$ is the covariance matrix, and
$\Sigma_{kl} := \langle\!\langle (\Omega_k - \mu_k)(\Omega_l -
\mu_l)\rangle\! \rangle$ for a Gaussian process $\vec\Omega$.

With no correlations between the stochastic processes $\omega_1(t),
\omega_2(t)$, and $\omega_3(t)$, the covariance matrix is diagonal:
$\Sigma= \mbox{diag} (2 \varsigma_1^2, 2 \varsigma_2^2, 2
\varsigma_3^2)$. Note that $\sqrt{2}\varsigma_k$ simply denotes the
standard deviation of the Gaussian random variable $\Omega_k$. The
preceding {\it commutativity relation} (\ref{eq:commutativity}) now
allows for a separation of the channel into its mixing dynamics and
its entangling dynamics: Any change with respect to the purity of the
system is due to the noise channel $\tilde D$, while any increase in
entanglement can only be evoked by (the interaction part of) the mean
Hamiltonian $H_\mu$. Of course, a decrease in entanglement might as
well be due to $\tilde D$. However, any gain in entanglement may only
be attributed to the unitary map $U_\mu$. This separation enables us
to calculate changes in purity and concurrence separately.

In order to describe the loss of purity of the system, we need to
account for the action of the noise channel, $\tilde D$, only,
\begin{eqnarray}
  \label{eq:Dtilde}
  \tilde D \star \rho &=& p_1 \rho + p_2 (\sigz^A \ten \sigz^B\; \rho
  \; \sigz^A\ten\sigz^B) \\
  &+& \! \! p_3 (\sigz^A\ten\id \; \rho \;\sigz^A\ten\id) + 
  p_4 (\id\ten\sigz^B \; \rho \; \id \ten \sigz^B),
  \nonumber 
\end{eqnarray}
where the probabilities $p_i$ with $\sum p_i=1$ are determined by the
variances of the stochastic processes (see Appendix
\ref{app:probabilities}). The purity for the final state is easily
obtained and equates to
\begin{eqnarray}
  \label{eq:purity}
  P(\tilde D \star \rho) &=& (p_1^2+p_2^2+p_3^2+p_4^2) \\
  &+&2\big(p_1p_3+p_2p_4 \big)\, |\expect{\sigz^A}|^2 \nonumber \\
  &+&2\big(p_1p_4+p_2p_3 \big)\, |\expect{\sigz^B}|^2 \nonumber \\
  &+&2\big(p_1p_2+p_3p_4 \big)\, |\expect{\sigz^A \ten \sigz^B}|^2.  \nonumber
\end{eqnarray}
Here and in the following we write $\expect{\,\cdot\,}$ for
expectation values with respect to the initial state
$\ket{\psi_0}$. While the equation for purity decay under RU phase
damping is valid irrespective of further details of the initial state,
the change in entanglement will be more sensitive to the initial
conditions of the two-qubit system. In order to calculate the
entanglement created or destroyed in the phase-damping process, we
thus need to make some further assumptions about the initial state.

\section{RU Single-Qubit Channels in the CP Diagram}
\label{sec:CPdiagram}

\begin{figure}[t]
  \includegraphics{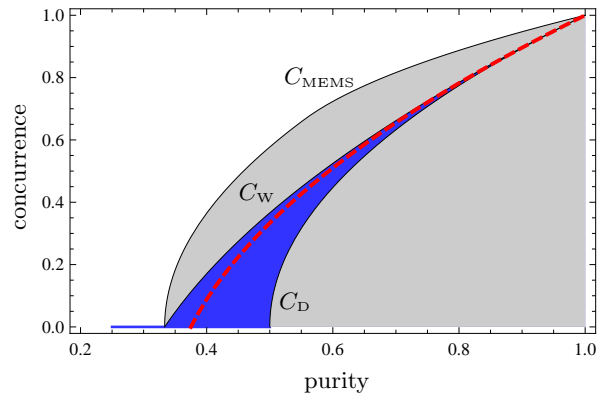}
  \caption{(Color online) The CP diagram. The light gray area shows
    the region of all physical states of two qubits, bounded by the
    maximally entangled mixed states ($\Cm$). The possible $C(P)$
    relations for a Bell state subject to single-sided unital dynamics
    or, equivalently, of all doubly chaotic states, is highlighted in
    blue (dark gray). This region is bounded by the curves valid for
    Werner states $\Cw$ and for Bell states under single-sided phase
    damping $\Cd$, respectively. The diagram also reveals the maxium
    $C(P)$ relation of doubly chaotic states with Bell rank $3$,
    $C_3(P)$ (dashed line). }
  \label{fig:figure1}
\end{figure}

In the analysis of RU phase damping we want to follow the example of
Ref.~\cite{Ziman2005}, where single-sided unital dynamics was analyzed
with respect to entanglement decay as a function of purity decay. The
entanglement for a two-qubit system can be easily computed in terms of
concurrence, $C(\rho) = \max\{0, \lambda_1 - \lambda_2 - \lambda_3 -
\lambda_4\}$, where $\lambda_1^2 \geq \lambda_2^2 \geq \lambda_3^2
\geq \lambda_4^2$ are the eigenvalues of the positive matrix $R=\rho
\, (\sigma_y \otimes \sigma_y)\, \rho^* \, (\sigma_y \otimes
\sigma_y)$ \cite{Wootters1998}. The purity $P$ is defined as the trace
over the squared density operator: $P(\rho)= \tr(\rho^2)$. Accessible
values of purity $P$ and concurrence $C$ are not independent of each
other. Rather, it has been shown that---depending on the entanglement
measure used---there exist states that maximize the respective
entanglement measure at a given mixedness, also called {\it maximally
  entangled mixed states} (MEMS, see \cite{Ziman2005} and references
therein). This fact is nicely illustrated in the {\it
  concurrence-vs-purity} diagram (CP diagram; see
Fig. \ref{fig:figure1}).

Aside from the border of all physical states given by $\Cm$, the
diagram reveals the accessible region (blue) for initial Bell states
under single-sided unital channels $\eps \otimes \id$, that is, unital
channels acting on either one of the subsystems, only. This accessible
region found is bounded by the curves $\Cw$ and $\Cd$ that are
realized for single-sided depolarization and single-sided phase
damping, respectively. The former is also valid for the so-called
Werner states \cite{Werner1989}, $\rho_W = p \ket{\psi_S} \bra{\psi_S}
+ (1-p) \id /4$, a convex mixture of the pure and maximally entangled
singlet state $\ket{\psi_S}= (\ket{01} - \ket{10})/ \sqrt 2$ and the
maximally disordered two-qubit state $\id/4$.

These borders will also play an important role in our findings. We
start by rederiving the results found in Ref.~\cite{Ziman2005} in an
alternative way. It is known that in case of single-qubit dynamics
there is a one-to-one correspondence between unital channels and
channels that are RU \cite{LandauStreater}. By applying the quantum
channel $\eps$ to one part of a maximally entangled bipartite state,
the so-called {\it Jamiolkowski isomorphism} \cite{ZyczkowskiBook}
introduces a duality between quantum channels defined on a Hilbert
space $\mathcal H$ and quantum states living on a Hilbert space
$\mathcal H \otimes \mathcal H$ of squared dimension. It follows
straight away that the Jamiolkowski state of a RU channel is doubly
chaotic \cite{LandauStreater}, that is, the state has maximally
disordered subsystems: $\tr_{1} \rho = \tr_{2} \rho = \mathds 1/d$
(here, $\tr_{i}$, $i=1,2$, refers to a partial trace over the first
and second subsystems, respectively, while $d = \dim \mathcal H$).  In
this vein, any single-qubit RU channel may be identified with a doubly
chaotic two-qubit state. Any such state may in turn be obtained by
applying a local unitary transformation $U \ten V$ to a Bell-diagonal
state \cite{Horodecki1996}, that is, a state diagonal in the basis of
Bell states $\ket{\psi^{\tiny\mbox{B}}_i}$:
\begin{eqnarray*}
  \rho = U \otimes V 
  \left(\sum_{i=1}^m p_i \ketbra{\psi^{\tiny \mbox{B}}_i}{\psi^{\tiny \mbox{B}}_i} \right) 
  U^\dagger\otimes V^\dagger
\end{eqnarray*}
with $\sum_i p_i = 1$. 

For such a state the accessible area in the CP diagram may be
determined easily (see Appendix \ref{app:BellDiagonal}).  As a general
lower bound we find the relation $C(P)=\sqrt{2P-1}=\Cd$, whereas the
upper bound depends on the Kraus rank $k$ of the corresponding qubit
channel $\eps$, that is, on the minimum number of Kraus operators
needed in (\ref{eq:kraus}).  For $k=2$ we have
\begin{eqnarray}
  \label{eq:C2ofP}
  C_2(P) &=& \Cd = \sqrt{2P-1},\quad P \geq \frac{1}{2},
\end{eqnarray}
again identical to the lower bound in the CP diagram, whereas for
$k=4$ we have
\begin{eqnarray}
  \label{eq:C4ofP}
  C_4(P) &=& \Cw = \frac{1}{2} \left(\sqrt{12\, P - 3} - 1\right), 
  \quad P \geq \frac{1}{3},
\end{eqnarray}
reproducing the upper bound. In addition, we get the relation
\begin{eqnarray}
  \label{eq:C3ofP}
  C_3(P) &=& \frac{1}{3} \left(\sqrt{24\, P - 8} - 1\right),
  \quad P \geq \frac{3}{8},
\end{eqnarray}
giving an upper bound of the allowed area for single-sided RU channels
of Kraus rank 3 (see Fig.~\ref{fig:figure1}, dashed line).
  
Thus, we arrive at the same results that were obtained by applying
single-sided unital channels to a pure and maximally entangled
two-qubit state \cite{Ziman2005}. We find that the accessible region
in the CP plane depends on the Kraus rank of the RU channel. It is
then quite obvious that the lower and upper border are given by
single-sided phase damping (Kraus rank 2) and depolarizing (Kraus rank
4), respectively. The various bounds in the CP diagram may be
attributed to a characteristic trait of doubly chaotic states. This is
also true for the behavior discovered in \cite{Torres2009}, where for
an initial Bell state of two noninteracting qubits with symmetric
coupling to a cavity with photon number $n \rightarrow \infty$ a
convergence to $C_3(P)$ was found. In this limit, the Bell state is
simply mapped to a doubly chaotic state with Bell rank 3, hence
explaining the asymptotic behavior.

For {\it random local unitary} (RLU) channels, that is, channels of the
form $\eps(\rho)=\sum_{r,s} p_{rs} (U_r \otimes V_s) \rho (U_r^\dagger
\otimes V_s^\dagger)$, with $\sum_{r,s} p_{rs} =1$ and unitary
$U_r,V_s$, it is easy to see that a Bell state is mapped to a doubly
chaotic state. Comparison with Eq.~(\ref{eq:Dtilde}) shows that the
action of the noise channel $\tilde D$ is exactly of this RLU
nature. Therefore, it is by no means surprising that the bounds play a
role in our findings. Note, however, that our studies cover not only
the case of maximally entangled initial states, but of pure initial
states in general. Based on our derivation it also follows immediately
that the accessible region for {\it bilocal} unital dynamics $\eps_1
\otimes \eps_2$---a question raised in Ref.~\cite{Ziman2005}---is in
fact identical to the area obtained for single-sided unital channels.

\section{Separable initial state}

In Sec. \ref{sec:RUPD} we discussed the possible separation of the RU
phase-damping channel, making it possible to separately calculate the
evolution of purity and entanglement. The general formula for purity
of the final state was already given, Eq.~(\ref{eq:purity}), whereas
in order to study entanglement, we need to make further assumptions
about the initial state. In this first section we take the initial
state to be separable. Due to the commutativity of the noise channel
$\tilde D$ and the unitary part $U_\mu$, we can start out by bringing
a pure separable state $\rho := \ket{\psi_0} \bra{\psi_0} :=
\ket{\phi_1} \bra{\phi_1} \ten \ket{\phi_2} \bra{\phi_2}$ into the
mixture $\tilde D\star \rho$, prior to applying $U_\mu$. For the final
state $\rho'$ we will then calculate the concurrence $C$. For generic
single-qubit states we can let
\begin{eqnarray*}
  \ket{\phi_i} := 
  \begin{pmatrix} \alpha_i \\
    \beta_i 
  \end{pmatrix},\; i=1,2, 
\end{eqnarray*}
with $\alpha_i, \beta_i \in \mathds C, |\alpha_i|^2+|\beta_i|^2
=1$. The calculation of the concurrence of $\rho'$ involves the
determination of the eigenvalues of the positive, non-Hermitian matrix
$R = \rho' (\sigma_y \otimes \sigma_y) \rho'^* (\sigma_y \otimes
\sigma_y)$. We find (see Appendix \ref{app:Rmat})
\begin{eqnarray}
  \label{eq:ConMittel}
  C(\rho^\prime)  &=&g(\ket{\psi_0}) \cdot f(\eps), 
\end{eqnarray}
where 
\begin{eqnarray}
  g(\ket{\psi_0}) &=&  4|\alpha_1| |\beta_1| |\alpha_2| |\beta_2| 
  = \sqrt{\mbox{var}(\sigma_z^A)}\sqrt{\mbox{var}(\sigma_z^B)}, \nonumber \\
  f(\eps)  &=& \mbox{max} \Big \{ 0, (p_1-p_2) \sin \mu_3 \nonumber \\ 
  && \quad - \sqrt{ (p_3+p_4)^2 \sin^2 \mu_3 + 
    4 p_3p_4 \cos^2 \mu_3 } \Big\}. \nonumber 
\end{eqnarray}
With $\mbox{var}(\sigma_z^i) = 1 - \expect{\sigma_z^i}^2$ we denote
the variances of $\sigma_z^i$ ($i=A,B$), which can be taken with
respect to the initial state $\ket{\psi_0}$.  The mean value $\mu_3 =
\average{\Omega_3}$, as well as the probabilities $p_i$, were
introduced in Sec.~\ref{sec:RUPD} and reflect mean value and variance
of the fluctuating fields (see also App.~\ref{app:probabilities}).

The concurrence thus allows for an evolution equation factorizing into
two functionals $f$ and $g$ that depend on the phase-damping channel
$\eps$ and the initial state of the qubits $\ket{\psi_0}$,
respectively.  The evolution equation accounts for the gain in
entanglement due to the unitary map $U_\mu$ through the function $f$,
while $g$ simply gives a scaling factor for the amount of entanglement
a certain initial state may achieve.

When trying to detect the accessible area within the CP diagram we
need to know the maximum of the concurrence with respect to the
entangling part of the channel. This maximum is given for $\sin \mu_3
=1$ or, equivalently, $\mu_3 \in \{(2k+1) \frac{\pi}{2} \;|\; k \in
\mathds Z \}$, which leads to the simple equation
\begin{eqnarray} 
  \label{eq:concurrence2}
  C_{\mbox{\tiny Max}} &=& \!\mbox{max} \Big\{ 0, g(\ket{\psi_0}) \;
  (p_1-p_2-p_3-p_4) \Big\} \nonumber \\
  &=& \!\mbox{max} \Big\{ 0, g(\ket{\psi_0}) \\
  && \quad \times \frac{1}{2} \Big[ e^{-\left(\varsigma_1^2+ \varsigma_2^2\right)} + 
  e^{-\left(\varsigma_1^2+ \varsigma_3^2\right)}+ 
  e^{-\left(\varsigma_2^2+ \varsigma_3^2\right)} - 1 \Big]\! \Big \}. \nonumber
\end{eqnarray}

\subsection{Pure two-qubit phase damping}

We want to discuss the results obtained so far in a setting where we
admit only pure two-qubit phase damping. By this we mean the situation
where the phase damping is caused by a fluctuation in the part of the
Hamiltonian describing the coupling of the two qubits only. Note that
this setting is also of relevance in the realization of a phase gate
in recent ion-trap experiments \cite{Monz2009}. The Hamiltonian then
consists of the single term $H (t) = \omega_3(t)
\sigz^A\ten\sigz^B$. Recall the definition of the variances $2
\varsigma_k^2 = \average{\Omega_k^2} - \average{\Omega_k}^2$
(cf. Sec.~\ref{sec:RUPD}).  We may therefore let the variances
$\varsigma_1= \varsigma_2=0$, and Eq.~(\ref{eq:concurrence2})
simplifies to
\begin{eqnarray}
  \Cmax (\rho') =  e^{- \varsigma_3^2} \cdot g(\ket{\psi_0}),
\end{eqnarray}
whereas the purity simplifies to
\begin{eqnarray}
  P(\rho') &=& \frac{1}{2} ( 1 + e^{-2 \varsigma_3^2}) + \\
  &&\quad \frac{1}{2} ( 1 - e^{-2 \varsigma_3^2}) 
  \cdot |\expect{\sigz^A}|^2 \cdot |\expect{\sigz^B}|^2. \nonumber
\end{eqnarray}
The simple form of both the maximal concurrence and the purity enables
us to give the former as a function of the latter:
\begin{eqnarray}
  \label{eq:CvonP-reine-zwei}
  \Cmax (P) &=& \sqrt{ \frac{\big( 1 - \expect{\sigma_z^A}^2 \big) \,
      \big( 1 - \expect{\sigma_z^B}^2 \big) }{1 - \expect{\sigma_z^A}^2 
      \expect{\sigma_z^B}^2 }} \times \nonumber \\
  && \quad \sqrt{ 2 P - 1 - \expect{\sigma_z^A}^2 
    \expect{\sigma_z^B}^2}. 
\end{eqnarray}
Note that for $\langle \sigma_z^A \rangle = \langle \sigma_z^B
\rangle=0$ we get $\Cmax (P)=\sqrt{2P-1}$, which is exactly the
equation obtained for the lower bound $\Cd$ shown in
Fig.~\ref{fig:figure1}.

\begin{figure}[t]
  \includegraphics{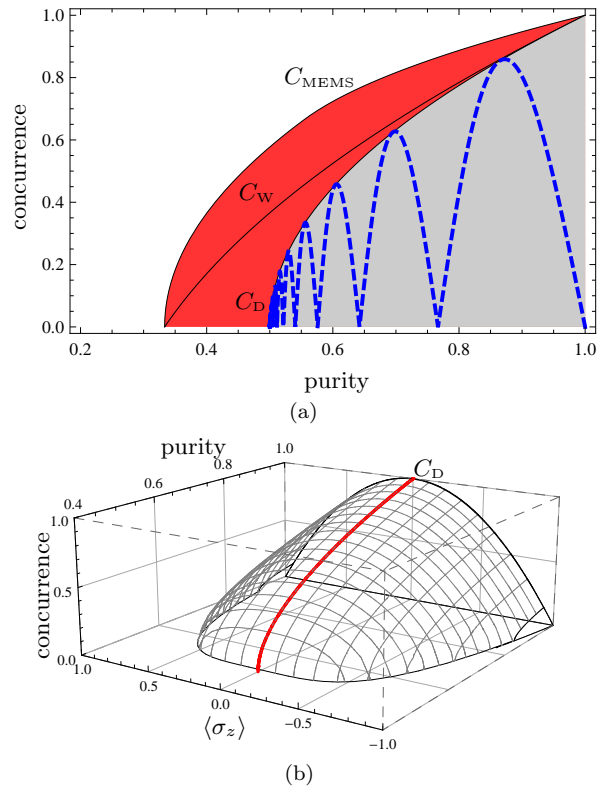}
  \caption{(Color online) (a) For a pure separable initial state
    subject to pure two-qubit phase damping, the maximally achievable
    concurrence is bounded from above by $\Cd$. The corresponding
    forbidden area is highlighted in red (dark gray). The blue
    (dashed) line represents the evolution under a one-parameter class
    of phase-damping channels (see text for details). (b) The
    accessible value of concurrence depends on expectation value
    $\langle \sigma_z^{(i)} \rangle$ evaluated for either one of the
    qubits $i=A,B$. Only for $\langle \sigma_z^A\rangle =
    \langle\sigma_z^B\rangle = 0$ may the relation $\Cd$ (red, thick
    line) be obtained.}
  \label{fig:figure2}
\end{figure}

In Fig.~\ref{fig:figure2} we show the CP diagram where we highlight
the excluded area. In order to show exemplary dynamics in the diagram
we let both mean and variances of the stochastic processes in
Eq.~(\ref{eq:hamil-zweiqubit}) depend quadratically on some ``time
parameter'' $t$. This introduces a one-parameter class of channels
$\eps_t$ mimicking continuous dynamics. A possible evolution is shown
as blue line.

Equation (\ref{eq:CvonP-reine-zwei}) relating maximal concurrence and
purity allows for a representation of the accessible region under pure
two-qubit phase damping (for an initially separable state). For this
we define $Z$ as the expectation value of the number of excited
states, that is, $Z:= 1 + \frac{\expect{\sigz^A}+
  \expect{\sigz^B}}{2}$.  When maximizing the function $\Cmax (P)$ for
fixed $P$ and $Z$, we get the requirement $\langle\sigz^A\rangle =
\langle\sigz^B\rangle =: \expect{\sigz}$. Inserted in
(\ref{eq:CvonP-reine-zwei}), this leads to
\begin{eqnarray}
  \label{eq:1}
  \Cmax (P,\expect{\sigz}) = \sqrt{\frac{1-\expect{\sigz}^2}
    {1+\expect{\sigz}^2} (2P-1-\expect{\sigz}^4)}. 
\end{eqnarray}
This relation can now be visualized in the
concurrence-purity-$\expect{\sigz}$ space [cf.~Fig.~\ref{fig:figure2}
(b)]. This representation was also used in \cite{McHugh2006}, where
the physically allowed region of a two-qubit system with respect to
concurrence, purity, and energy was studied.

\subsection{General uncorrelated phase damping}

Next we want to consider the general case of uncorrelated phase
damping. Remember that by uncorrelated we mean a diagonal covariance
matrix $\Sigma$. Here, we use the full dynamics, that is, the part of the
Hamiltonian describing pure two-qubit dephasing as well as the parts
acting locally on both qubits.

As in the preceding section, the concurrence attains its maximum for
$\langle \sigma_z^A \rangle = \langle \sigma_z^B \rangle=0$, leading
to the identities
\begin{eqnarray}
  \label{eq:Cuncorr}
  C &=& \frac{1}{2} \left[ e^{-\left(\varsigma_1^2+\varsigma_2^2\right)} 
    + e^{-\left(\varsigma_1^2+\varsigma_3^2\right)}
    + e^{-\left(\varsigma_2^2+\varsigma_3^2\right)} - 1 \right] \\
  \label{eq:Puncorr}
  P &=& \frac{1}{4} \left[ e^{-2\left(\varsigma_1^2+\varsigma_2^2\right)}
    + e^{-2\left(\varsigma_1^2+\varsigma_3^2\right)}
    + e^{-2\left(\varsigma_2^2+\varsigma_3^2\right)} + 1 \right]. 
\end{eqnarray}
Note that for identical variances $\varsigma_1 = \varsigma_2 =
\varsigma_3 =: \varsigma$, this implies $ C = \frac{1}{2} \big(3 e^{-2
  \varsigma^2} - 1 \big)$ and $P = \frac{1}{4} \big(3 e^{-4
  \varsigma^2} + 1 \big)$, which is easily transformed into the
relation
\begin{eqnarray}
  \label{eq:werner}
  C(P) = \frac{1}{2} \left( \sqrt{ 12\, P-3} - 1 \right), \quad P \geq \frac{1}{3}, 
\end{eqnarray}
which coincides with the relation obtained for the Werner states,
$\Cw$.

\begin{figure}[t]
  \includegraphics{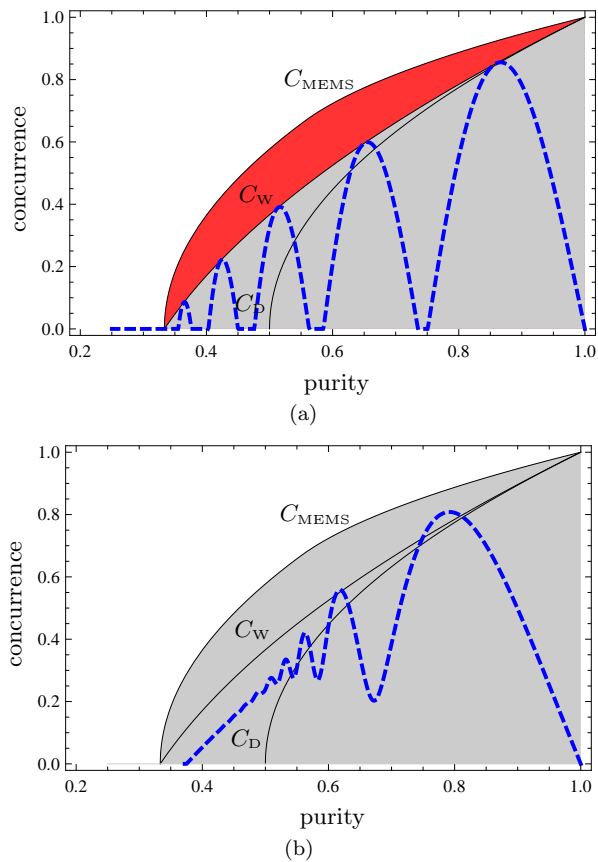}
  \caption{(Color online) (a) In case of general uncorrelated phase
    damping, the maximally achievable concurrence is identical to the
    value given by the Werner states, $\Cw$. As before, the forbidden
    area is highlighted in red (dark gray). The role of correlations
    may be observed in (b), where both bounds are violated. In both
    diagrams, an exemplary evolution under a one-parameter class of
    phase-damping channels is shown (dashed blue lines). }
  \label{fig:figure3}
\end{figure}

Furthermore, from Eqs.~(\ref{eq:Cuncorr}) and (\ref{eq:Puncorr}) we
conclude that $C^2 + C + 1 \leq 3P$, so that the case of all variances
being equal already yields the maximal relation of concurrence as a
function of purity. We conclude that $\Cmax (P) = (\sqrt{12P-3} -
1)/2$ in case of uncorrelated phase damping. For pure, separable
states subject to RU phase damping, we can therefore identify a
forbidden area in the CP plane, the border of which is given by the
Werner states $\Cw$. In Fig.~\ref{fig:figure3} (a) we highlight the
excluded area for general, uncorrelated phase damping.

Note that these results are only true for uncorrelated phase damping.
In Fig.~\ref{fig:figure3} (b) we show an example of correlated phase
damping, that is, with nondiagonal covariance matrix $\Sigma$ (see
Sec.~\ref{sec:RUPD}). This accentuates the fact that the borders in
the CP diagram are valid only in the uncorrelated case. We ascribe
this violation of the bounds in case of correlated phase damping to
the fact that the irreversible part of the channel $\tilde D$ is in
general no longer RLU (see Sec.~\ref{sec:CPdiagram}).

\section{General Pure Initial  State}

In this section we will analyze the evolution of a general pure
initial state under phase damping, that is, we drop the restriction of
separability for the initial state. For our purposes, that is, for
entanglement and purity evolution under phase damping, the most
general (pure) intial state reads
\begin{eqnarray*}
  \ket{\psi_0} &=& \big( e^{-\frac{i}{2} \theta_1 \sigma_x} 
  \ten e^{-\frac{i}{2} \theta_2 \sigma_x}\big) \times  \\
  && \quad \quad  \left( \sqrt{a} \ket{00} + e^{-i \chi} \sqrt{1-a} \ket{11}\right), 
  \nonumber
\end{eqnarray*}
leaving only four parameters $(a, \chi, \theta_1, \theta_2)$ to fully
characterize the initial state of the two-qubit system (see Appendix
\ref{app:GPIS}).

While Eq.~(\ref{eq:purity}), giving the purity of the final state, is
still valid, we need to estimate the decay of the initial concurrence
when the state is subject to the mixing part of the phase damping
channel, $\tilde D$. In close analogy to the method described in
Appendix \ref{app:Rmat}, we are again able to estimate the concurrence
analytically:
\begin{eqnarray}
  C(\rho^\prime)&=& \mbox{Max} \Big\{ 0, \nonumber \\
  && \!\!\!\!\!\!\! \sqrt{(p_1^2+p_2^2) |s|^2+
    2p_1p_2|r|^2-2p_1p_2|s^2-r^2|} - \nonumber \\
  && \!\!\!\!\!\!\! \sqrt{(p_3^2+p_4^2) |s|^2 + 2 p_3p_4 |r|^2 + 
    2 p_3p_4 |s^2-r^2| } \Big \}, \nonumber
\end{eqnarray}
where we have defined $s := \bra{\psi_0} \Sigma_y \ket{\psi_0^*} $ and
$r := \bra{\psi_0} \Sigma_z \Sigma_y \ket{\psi_0^*}$. Here we have
also used the abbreviations $\Sigma_i := \sigma_i \otimes \sigma_i$
($i=x,y,z$).

Together with the purity (\ref{eq:purity}) we can now examine the
possible $C(P)$ combinations in more detail. In order to get
illustrative results, we again have to put some further constraints on
the dynamics.

\subsection{Single-sided phase damping}

For a single-sided phase-damping channel, that is, acting on either
one of the qubits, we can generalize the results obtained for
maximally entangled states in Ref.~\cite{Ziman2005} to the case of
pure initial states with arbitrary entanglement. Note that for this
scenario the channel is simply given by $\rho' = (1-q) \rho + q
(\sigma_z^A\ten \id) \rho (\sigma_z^A\ten \id)$.

\begin{figure}[t]
  \includegraphics{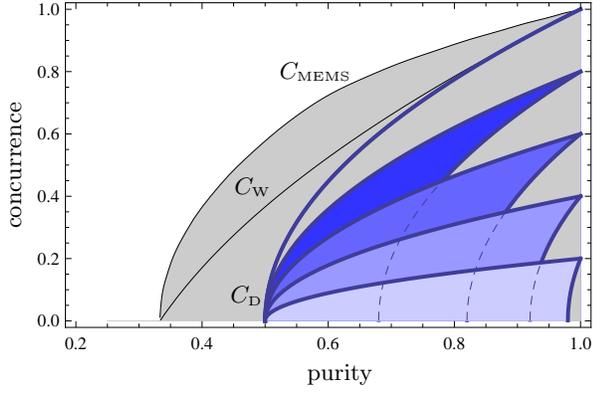}
  \caption{(Color online) Accessible region for a single-sided phase
    damping channel for states of varying initial entanglement
    $C_{\scriptstyle 0}=0.2\; (0.4,\; 0.6,\; 0.8,\; 1)$. For
    $C_{\scriptstyle 0} = 1$, the lower and upper bounds are equal,
    given by the known curve $\Cd$. }
  \label{fig:figure4}
\end{figure}

For single-sided phase damping it is easy to arrive at the evolution
equation of entanglement, which is given by
\begin{eqnarray*}
  C(\rho') = f(\eps) \cdot C_0,
\end{eqnarray*}
where $f(\eps) = 1-2q$. The state-dependent behavior within the CP
diagram thus stems from the evolution of purity, which is given by
\begin{eqnarray*}
  P(\rho') = 1 - 2q(1-q) \left[1 - |\expect{\sigz^A}|^2\right], 
\end{eqnarray*}
where $|\expect{\sigz^A}|^2 = \cos^2 \theta_1 (2a -1)^2$. 

We find that the $C(P)$ relation for given initial entanglement now
only depends on the angle $\theta_1$ . The extremal cases may be
identified as
\begin{eqnarray*}
  C(P) = 
  \left\{ 
    \begin{array}{cc} 
      \sqrt{C_0^2 (2P-1)} & \mbox{if } \cos^2\theta_1 =0 \\
      \sqrt{C_0^2+ 2(P-1)} & \mbox{if } \cos^2\theta_1=1 
    \end{array} 
  \right.
\end{eqnarray*}
where $C_0$ denotes the concurrence of the initial state
$\ket{\psi_0}$. For all other cases ($0 \leq \cos^2 \theta_1 \leq 1$),
one can see that the relation $C(P)$ is bounded by these two
curves. For maximally entangled initial states the two functions
coincide, again giving the by-now well-known relation $C(P)=
\sqrt{2P-1}=\Cd$ \cite{Ziman2005}. For nonextremal initial
entanglement $0 < C_0 < 1$, however, the two curves part, bounding the
accessible region from above and below. For states of arbitrary
initial entanglement subject to single-sided phase damping we conclude
that there exists a maximal value of purity, $P_< = 1 - C_0^2/2$,
above which all states are still entangled. Again, the results are
easily illustrated using the CP diagram (see Fig.~\ref{fig:figure4}).

\subsection{Pure two-qubit phase damping}

In case of pure two-qubit phase damping, the noise channel is of the
simple form $\rho'=(1-q)\rho + q (\sigz^A\ten \sigz^B) \rho
(\sigz^A\ten \sigz^B)$. Again we are able to give the evolution
equation of entanglement,
\begin{eqnarray}
  \label{eq:pure2Qu}
  C^2(\rho') &=& C_0^2 - 2q(1-q) \left[ |r|^2-|s|^2-|r^2-s^2| \right] \nonumber\\
  &=:& C_0^2 -  f(\eps)\cdot g(\ket{\psi_0}), 
\end{eqnarray}
where now $f(\eps) = 2q(1-q)$ and $g(\ket{\psi_0}) =
|r|^2-|s|^2-|r^2-s^2|$.  Entanglement depends on the initial state of
the two qubits through the functional $g$. Note that here the
evolution equation no longer follows the factorization law, as in the
case of single-sided channels. Yet the evolution is fully determined
through the functionals $f$ and $g$. The purity equates to
\begin{eqnarray}
  \label{eq:pure2QuPur}
  P(\rho') = 1- 2q(1-q) \left[ 1- |\expect{\sigz^A \ten \sigz^B}|^2 \right]
\end{eqnarray}
with $|\expect{\sigz^A \ten \sigz^B}|^2 = \big[\cos \theta_1 \cos \theta_2
- \sin \theta_1 \sin \theta_2 \cdot \cos \chi \cdot 2\sqrt{a
  (1-a)}\big]^2$.

\begin{figure}[t]
  \includegraphics{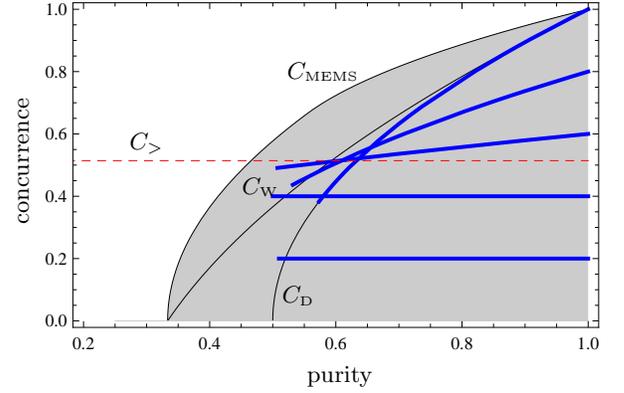}
  \caption{(Color online) For pure two-qubit phase damping we can
    identify states with robust entanglement. The plot is obtained for
    an initial state parameterized by $\chi=0,\, \theta_1=3 \pi/16,\,
    \theta_2 = \pi/8$ (see text), leading to robust entanglement for
    $C_0 \leq C_> \approx 0.5142$ (dashed red line).  The blue (thick)
    lines show exemplary evolution of pure states with initial
    concurrence $C_0 = 1.0 \; (0.8,\; 0.6,\; 0.4,\; 0.2)$.}
  \label{fig:figure5}
\end{figure}

Eq.~(\ref{eq:pure2Qu}) highlights the possibility of robust
entanglement under pure two-qubit phase damping. It is easy to see
that for $g (\ket{\psi_0}) = 0$ concurrence remains constant, whereas
purity [Eq.~(\ref{eq:pure2QuPur})] may, in general, still decline. We
find that this is the case for states with $\chi \in \{ k \pi | k \in
\mathds Z \}$, only. The square of the concurrence then becomes
\begin{eqnarray*}
  C^2(\rho')  =  
  \left\{ 
    \begin{array}{l@{\quad}l}
      s^2  - 4 q(1-q) [ C_0^2 - r^2 ] & \mbox{if } s^2 > r^2 \\
      s^2 & \mbox{if } s^2 \leq r^2. 
    \end{array}
  \right.
\end{eqnarray*}

The necessary condition for robust entanglement under pure two-qubit
phase damping is thus given by $s^2\leq r^2$ and $\chi = k \pi$. To be
specific, the necessary conditions for $\chi = k 2 \pi$ [$\chi =
(2k+1) \pi$] are given by
\begin{eqnarray*}
  & s^2 \leq r^2 &  \\
  & \Leftrightarrow & \\
  & C_0 \leq C_> := \mbox{max}\left\{ [-] \frac{\sin \theta_1 \sin\theta_2}
    {\cos\theta_1 \cos \theta_2 + 1}, [-] \frac{\sin \theta_1 \sin\theta_2}
    {\cos\theta_1 \cos \theta_2 - 1} \right\},
\end{eqnarray*}
hereby defining the upper bound $C_>$, such that states with initial
concurrence $C_0 < C_>$ exhibit robust entanglement. Note that for
$\cos\theta_1 \cos\theta_2 = \pm 1$, $s^2=r^2$ and, thus, constant
concurrence $C$ follows immediately. In Fig.~\ref{fig:figure5} we show
an example of robust entanglement. It can be clearly observed that for
initial concurrence smaller than $C_>$ and despite the loss of purity,
the entanglement is persistent. Beyond the observation of robust
entanglement we also see that, while the map of the maximally
entangled state follows the line $\Cd$, as would be predicted (the
Kraus rank of pure two-qubit phase damping is 2, see Sec.~III), for
states with initial concurrence $C_0<1$, the restricted area no longer
plays a role.

\section{Conclusions}

We have studied a two-qubit quantum system subject to phase damping
caused by fluctuating fields. Irrespective of possible correlations
between the individual constituents of the introduced stochastic
Hamiltonian, the quantum channel was shown to allow a separation into
two parts, representing reversible and irreversible parts of the
dynamics, respectively. The separation enabled us to estimate the
evolution of the system's entanglement, as well as purity,
analytically.  When combined, the results allow us to identify
exclusive regions within the concurrence-purity plane, suggesting a
connection between RLU dynamics on a Bell state and phase-damping
processes on a separable initial state. These {\it forbidden places},
however, do not play a role in the evolution of pure states of
arbitrary initial entanglement $0<C_0<1$.

Our approach enables us to generalize results obtained in
Ref.~\cite{Ziman2005}, where the action of single-sided unital
channels on Bell states was studied. Here we are able to analyze the
effect of single-qubit phase damping on pure two-qubit states of
arbitrary initial entanglement. For a certain class of phase damping
we have identified necessary conditions leading to a robustness of the
entanglement present in the two-qubit system. This robustness is
closely related to the entanglement of the initial state; more
precisely, depending on the initial state, we can give an upper bound
$C_>$ such that for all states with concurrence $C_0<C_>$ the
entanglement is robust. The crucial conclusion to be drawn from these
findings is that in order to preserve entanglement, sometimes {\it
  less is more}.

\acknowledgments 
  The authors thank 
  Gernot Alber, 
  Hartmut H\"{a}ffner,
  Carlos Pineda,
  Stephan Rietzler, 
  Thomas Seligman, 
  and Lars W\"urflinger for fruitful 
  discussions and hints. W.T.S. is
  grateful to the Centro Internacional de Ciencias in Cuernavaca,
  Mexico, where part of this work took shape. J.H. acknowledges
  support from the International Max Planck Research School (IMPRS)
  Dresden.

\appendix

\section{}
\label{app:RUchannels}
In Sec.~\ref{sec:RUPD} we present the decomposition
(\ref{eq:commutativity}) of the phase-damping channel into unitary
part and noise channel. In order to see this, recall the definition of
the RU channel
\begin{eqnarray*}
  \rho \mapsto \rho' = \average{U(t) \rho U^\dagger (t)}. 
\end{eqnarray*}
The unitary map is given by $U(t) = e^{-i \int_0^t d\tau H(\tau)}$, where
in our model
\begin{eqnarray*}
  H(t) = 
  \omega_1(t) \sigz^A\ten\id + \omega_2(t) \id \ten\sigz^B 
  + \omega_3(t) \sigz^A\ten \sigz^B. 
\end{eqnarray*}
A simple calculation gives the diagonal map
\begin{eqnarray}
  \label{app:eq:diagonal}
  \rho'_{mn} &=& \average{ e^{i \sum_{k=1}^3 c_k^{mn} \Omega_k}}\rho_{mn} \nonumber\\
  &=& e^{i \sum_{k=1}^3 c_k^{mn} \average{\Omega_k}} 
  \average{e^{i \sum_{k=1}^3 c_k^{mn} \left(\Omega_k-\average{\Omega_k}\right)}} \rho_{mn} 
  \nonumber\\
  &=:& D^\mu_{mn} \tilde D_{mn} \rho_{mn}, 
\end{eqnarray}
where we have used the compact notation $c_k^{mn} := (\vec e_m-\vec
e_n)_k $ with vectors $\vec e_1 = (0,0,0), \vec e_2 = (0,1,1),\vec
e_3=(1,0,1), \vec e_4 =(1,1,0)$, as well as the abbreviations
$D^\mu_{mn} = e^{i \sum_{k=1}^3 c_k^{mn} \average{\Omega_k}}$ and $
\tilde D_{mn} = \average{e^{i \sum_{k=1}^3 c_k^{mn} \left(\Omega_k -
      \average{\Omega_k}\right)}}$. The diagonal map $D^\mu_{mn}$ may
be rewritten in terms of a diagonal unitary transformation $U_\mu =
e^{-i H_\mu t}$ such that $D^\mu_{mn} = (U_\mu)^{}_{mm}
(U_\mu)^*_{nn}$, where the Hamiltonian $H_\mu$ is determined by the
mean values $\average{\Omega_k}=:\mu_k$ :
\begin{eqnarray*}
  H_\mu := \mu_1 \sigz^A\ten\id + \mu_2 \id\ten\sigz^B + \mu_3
  \sigz^A\ten\sigz^B. 
\end{eqnarray*}
Commutativity of unitary part $U_\mu$ and noise channel $\tilde D$ is
then a direct consequence of the diagonality (\ref{app:eq:diagonal})
of the two maps. Formally, we may thus write
\begin{eqnarray*}
  \rho' &=& U_\mu ( \tilde D\star \rho) U_\mu^\dagger \\
  &=& D^\mu \star (\tilde D \star \rho) \\
  &=& \tilde D \star (D^\mu \star \rho) \\
  &=& \tilde D \star \left ( U_\mu \rho U_\mu^\dagger \right). 
\end{eqnarray*}
\section{}
\label{app:probabilities}
The probabilities used in Sec. \ref{sec:RUPD} in terms of the
variances $2 \varsigma_k^2 = \average{\Omega_k^2} -
\average{\Omega_k}^2$ are given as follows:
\begin{widetext}
  \begin{eqnarray*}
    p_1 &=& \frac{1}{8} 
    \left\{
      \left(1+e^{-\varsigma_1^2}\right) \left(1+e^{-\varsigma_2^2}\right) 
      \left(1+e^{-\varsigma_3^2}\right)+
      \left(1-e^{-\varsigma_1^2}\right) \left(1-e^{-\varsigma_2^2}\right) 
      \left(1-e^{-\varsigma_3^2}\right) 
    \right\} \\
    p_2 &=& \frac{1}{8} 
    \left\{
      \left(1+e^{-\varsigma_1^2}\right) \left(1+e^{-\varsigma_2^2}\right) 
      \left(1-e^{-\varsigma_3^2}\right)+
      \left(1-e^{-\varsigma_1^2}\right) \left(1-e^{-\varsigma_2^2}\right) 
      \left(1+e^{-\varsigma_3^2}\right) 
    \right\} \\
    p_3 &=& \frac{1}{8} 
    \left\{
      \left(1+e^{-\varsigma_1^2}\right) \left(1-e^{-\varsigma_2^2}\right) 
      \left(1-e^{-\varsigma_3^2}\right)+
      \left(1-e^{-\varsigma_1^2}\right) \left(1+e^{-\varsigma_2^2}\right) 
      \left(1+e^{-\varsigma_3^2}\right) 
    \right\} \\
    p_4 &=& \frac{1}{8} 
    \left\{
      \left(1+e^{-\varsigma_1^2}\right) \left(1-e^{-\varsigma_2^2}\right) 
      \left(1+e^{-\varsigma_3^2}\right)+
      \left(1-e^{-\varsigma_1^2}\right) \left(1+e^{-\varsigma_2^2}\right) 
      \left(1-e^{-\varsigma_3^2}\right) 
    \right\}.
  \end{eqnarray*}
\end{widetext}
\section{}
\label{app:BellDiagonal}
In Sec.~\ref{sec:CPdiagram} we argue that any unital single-qubit
channel may be identified with a doubly chaotic two-qubit state. A
given doubly chaotic state of two qubits may in turn be obtained by
applying a local unitary transformation $U\otimes V$ to a
Bell-diagonal state \cite{Horodecki1996}:
\begin{eqnarray*}
  \rho = U \otimes V 
  \left(\sum_{i=1}^m p_i \ketbra{\psi^{\tiny \mbox{B}}_i}{\psi^{\tiny \mbox{B}}_i} \right) 
  U^\dagger\otimes V^\dagger
\end{eqnarray*}
with probabilities $p_i$, $\{ \ket{ \Psi_i^{\tiny\mbox{B}}}
\}_{i=1,...,4}$ denotes the basis of Bell states. By rearrangement of
the Bell basis let $p_1\geq p_2\geq p_3\geq p_4$, so that $m$ denotes
the {\it Bell rank} of $\rho$, that is, the minimum number of Bell
states needed to represent the state $\rho$. Note that the Bell rank
is equal to the so-called {\it Kraus rank} of the channel, which gives
the minimum number of terms in the Kraus representation
(\ref{eq:kraus}). Using the Bell-diagonal representation, both
concurrence and purity are of a very simple form:
\begin{eqnarray}
  \label{eq:purbell}
  C(\rho) = \max \{ 0 , 2 p_1 - 1 \}, \quad \quad
  P(\rho) = \sum_{i=1}^m  p_i^2. 
\end{eqnarray}

The maximum relation $C(P)$ now depends on the Bell rank $m$. It may
be obtained by minimizing the purity for given concurrence. Let $p_1=:
1-q$, then from Eqs.~(\ref{eq:purbell}) we can conclude that the
purity is minimal if the remaining weights $p_i$ with $i > 1$ are
equal, $p_i = q/(m-1)$. We thus obtain
\begin{eqnarray*}
  C(\rho) &=& \max \{ 0 , 1-2q \}, \\
  P(\rho) &\geq& 1 - 2q \left( 1 - q \frac{m}{m-1} \right), 
\end{eqnarray*}
and we can immediately give an upper bound for the concurrence as a
function of purity: $C(P) \leq 2 \frac{m-1}{m} \sqrt{\frac{mP-1}{m-1}}
- \frac{m-2}{m} =: C_m(P)$. Inserting the nontrivial Bell ranks $m =
2,3,4$, we get the upper bounds of the accessible $C(P)$ relations for
single-sided RU channels with corresponding Kraus ranks. Moreover, we
see that for arbitrary Kraus rank $k$, $C(P)$ is bounded from below by
$C_2(P)$.
\section{}
\label{app:Rmat}
The calculation of the concurrence involves the determination of the
eigenvalues of the positive, non-Hermitian matrix
\begin{widetext}
  \begin{eqnarray*}
    R &=& \rho^{\prime} \; \tilde{\rho}^{\prime} \nonumber \\
    &=& 4 \alpha_1^* \beta_1^* \alpha_2^* \beta_2^* \times \\
    &&\Big[ \; 
    e^{-i \boldsymbol{\tilde \varphi} } \ket{\psi_0}
    \Big\{ 
    - i p_1^2 \sin \mu_3 \bra{\psi_0^*}\Sigma_y e^{-i \boldsymbol{\tilde \varphi} } 
    - p_1p_2 \cos \mu_3 \bra{\psi_0^*}\Sigma_y \Sigma_z e^{-i \boldsymbol{\tilde \varphi} } 
    \Big\} \nonumber\\ && \; + 
    e^{-i \boldsymbol{\tilde \varphi} } \Sigma_z \ket{\psi_0}
    \Big\{ 
    - p_1p_2 \cos \mu_3 \bra{\psi_0^*}\Sigma_y e^{-i \boldsymbol{\tilde \varphi} } 
    - i p_2^2 \sin \mu_3 \bra{\psi_0^*}\Sigma_y \Sigma_z e^{-i \boldsymbol{\tilde \varphi} }
    \Big\} \nonumber\\ && \; + 
    e^{-i \boldsymbol{\tilde \varphi} } \sigz \ten \id \ket{\psi_0} 
    \Big\{
    i p_3^2 \sin \mu_3 \bra{\psi_0^*}\Sigma_y (\sigz \ten \id) 
    e^{-i \boldsymbol{\tilde \varphi} } 
    + p_3p_4 \cos \mu_3 \bra{\psi_0^*}\Sigma_y (\id \ten \sigz) 
    e^{-i \boldsymbol{\tilde \varphi} }
    \Big\} \nonumber\\ && \; + 
    e^{-i \boldsymbol{\tilde \varphi} } \id \ten \sigz \ket{\psi_0} 
    \Big\{ 
    p_3p_4 \cos \mu_3 \bra{\psi_0^*}\Sigma_y (\sigz \ten \id) 
    e^{-i \boldsymbol{\tilde \varphi} } 
    + i p_4^2 \sin \mu_3 \bra{\psi_0^*}\Sigma_y (\id \ten \sigz) 
    e^{-i \boldsymbol{\tilde \varphi} }
    \Big\}  
    \Big], \nonumber
  \end{eqnarray*}
\end{widetext}
where we have defined $\Sigma_i := \sigma_i \ten \sigma_i$ ($i=x,y,z$)
and $\boldsymbol{\tilde \varphi} := \mu_3 \Sigma_z/2$.

We find the existence of two $R$-invariant linear subspaces $S, \tilde
S \subset \mathds C^4$, spanned by the vectors $\{e^{-i
  \boldsymbol{\tilde \varphi} }| \psi_0 \rangle , e^{-i
  \boldsymbol{\tilde \varphi} } \Sigma_z |\psi_0 \rangle \} =:
\{\nu_1,\nu_2\}$ and $\{e^{-i \boldsymbol{\tilde \varphi} } \sigz \ten
\id | \psi_0 \rangle ,e^{-i \boldsymbol{\tilde \varphi} } \id \ten
\sigz | \psi_0 \rangle \}=: \{\tilde\nu_1,\tilde\nu_2\}$,
respectively. For the determination of the eigenvalues of $R$ we thus
have to find a basis of orthogonal eigenvectors of spaces $S$ and
$\tilde S$. For instance, for eigenvalues $\lambda_+^2,\lambda_-^2$ in
$S$, we need to consider
\begin{eqnarray*}
  R ( \nu_1 + k \nu_2) &=& 
  a \nu_1 + b \nu_2 + k \left( c \nu_1 + d \nu_2 \right) \nonumber \\
  &\stackrel{!}{=}& \lambda^2_{\pm} \left( \nu_1 + k \nu_2 \right)
\end{eqnarray*}
where $a,b,c,d,k \in \mathds C$, leading to the quadratic equations
$\lambda^2_{\pm} = \frac{a + d}{2} \pm \sqrt{ \frac{(a-d)^2}{4} +
  bc}$. Note that in the formula for the concurrence there is a sum of
the square roots of the eigenvalues of the matrix $R$. Therefore the
relation $\left( \lambda_{+} \pm \lambda_{-}\right)^2 = (a + d) \pm 2
\sqrt{ad-bc}$ will prove to be quite useful. For the second subspace
$\tilde S$, the situation is analogous and the eigenvalues of $R$ are
thus given by the set $\{\lambda^2_{+}, \lambda^2_{-},
\tilde\lambda^2_{+}, \tilde\lambda^2_{-} \}$. The concurrence is
eventually given by
\begin{eqnarray*}
  C(\rho^\prime) &=& 
  \mbox{Max} \Big \{ 0,  4|\alpha_1| |\beta_1| |\alpha_2| |\beta_2| 
  \left(\lambda_+ - \lambda_- - \tilde\lambda_+ - \tilde\lambda_-\right) 
  \!\! \Big\}  \\
  &=&g(\ket{\psi_0}) \cdot f(\eps), 
\end{eqnarray*}
where we define
\begin{eqnarray}
  f(\eps) &:=&  \mbox{Max} \Big \{ 0, \left(\lambda_+ - \lambda_-
    - \tilde\lambda_+ - \tilde\lambda_-\right) \Big\} \nonumber \\
  &=& \mbox{Max} \Big \{ 0, (p_1-p_2) \sin \mu_3 - \nonumber \\ 
  && \quad \sqrt{ (p_3+p_4)^2 \sin^2 \mu_3 + 
    4 p_3p_4 \cos^2 \mu_3 } \Big\} \nonumber
\end{eqnarray}
and
\begin{eqnarray}
  g(\ket{\psi_0}) &:=&  4|\alpha_1| |\beta_1| |\alpha_2| |\beta_2| \nonumber.
\end{eqnarray}

\section{}
\label{app:GPIS}
By use of the Schmidt decomposition, any pure two-qubit state may be
written in the form $\ket{\psi} = U_1 \ten U_2 \left( \sqrt{a}
  \ket{00} + \sqrt{1-a} \ket{11} \right)$, where $U_1,U_2 \in
\mathcal{SU}(2)$ \cite{NielsenChuang}. Note that the concurrence
depends on the single parameter $a$ only and equates to $C(\ket{\psi})
= 2 \sqrt{a(1-a)}$. Using the Euler angles ($\varphi_i, \theta_i,
\chi_i$) the unitary rotations may be written in the form $U_i =
e^{-\frac{i}{2} \varphi_i \sigz} e^{-\frac{i}{2} \theta_i \sigma_x}
e^{-\frac{i}{2} \chi_i \sigz}$, $i=1,2$. Due to diagonality of both
the phase-damping channel and the last rotation, $e^{-\frac{i}{2}
  \varphi_1 \sigz}\ten\ e^{-\frac{i}{2} \varphi_2 \sigz}$, we may
reverse their order (diagonal operators commute). When interested in
entanglement and purity only, the invariance of concurrence under
local unitaries, as well as the cyclic invariance of the trace
operation, then make it possible to completely disregard the last
rotation.  The first rotation $e^{-\frac{i}{2} \chi_1 \sigz}\ten\
e^{-\frac{i}{2} \chi_2 \sigz}$ simply translates into a relative phase
$\chi := \chi_1 + \chi_2$.  For an analysis of entanglement and purity
of a pure state under phase damping it is thus sufficient to study
states of the form
\begin{eqnarray*}
  \ket{\psi} &=& \big( e^{-\frac{i}{2} \theta_1 \sigma_x} 
  \ten e^{-\frac{i}{2} \theta_2 \sigma_x}\big)   \\
  && \quad \quad  \times \left( \sqrt{a} \ket{00} + e^{-i \chi} \sqrt{1-a} \ket{11}\right).
\end{eqnarray*}

\bibliography{apsrev}{}

\end{document}